\documentclass[aps,prb,twocolumn,groupedaddress,floats,showpacs]{revtex4}
\usepackage{latexsym}
\usepackage{dcolumn}
\usepackage[dvips]{graphicx}
\usepackage{amssymb}
\usepackage{graphics}
\usepackage{amsmath}
\usepackage{epsf}

\newcommand{\vk}{{\bf k}}
\newcommand{\tat}{\tau_t}
\newcommand{\tas}{\tau_s}
\newcommand{\tr}{\tau_t/\tau_s}

\begin{document}
\title{Single particle relaxation time versus transport scattering
  time in a 2D graphene layer}
\author{E. H. Hwang and S. Das  Sarma} 
\affiliation{Condensed Matter Theory Center, Department of 
        Physics, University of Maryland, College Park, MD 20742-4111}
\date{\today}
\begin{abstract}
We theoretically calculate and compare the single-particle relaxation
time ($\tau_s$) defining quantum level broadening and the transport
scattering time ($\tau_t$) defining Drude conductivity in 2D graphene
layers in the presence of screened charged impurities scattering and
short-range defect scattering. We find that the ratio $\tau_t/\tau_s$
increases strongly with increasing $k_F z_i$ and $\kappa$ where $k_F$,
$z_i$, and $\kappa$ are respectively the Fermi wave vector, the
separation of the substrate charged impurities from the graphene
layer, and the background lattice dielectric constant. A critical
quantitative comparison of the $\tau_t/\tau_s$ results for graphene
with the corresponding modulation-doped semiconductor structures is
provided, showing significant differences between these two 2D carrier
systems.
\end{abstract}
\maketitle

\section{introduction and background}
\label{Section:Introduction}

In an impure disordered conductor (either a metal or a doped
semiconductor) scattering by static impurities and defects, in
general, leads to two distinct momentum relaxation times; the
scattering life time or the transport relaxation time (denoted by
$\tau_t$ in this paper, but often just called $\tau$) and the quantum
lifetime or the single-particle relaxation time (denoted by $\tau_s$
in this paper, but often also called $\tau_q$). Although $\tat$ and $\tas$
both arise from impurity scattering in the metallic regime,
they are in general distinct and unique with no direct analytical
relationship connecting them, except for the simple (and often
unrealistic) model of completely isotropic $s$-wave zero-range
impurity scattering when they are equal. In particular, $\tat$
determines the conductivity, 
$\sigma = ne\mu \propto \tat$ where $n$ is the carrier density and
$\mu$ the mobility, whereas $\tas$ determines the quantum level
broadening, $\Gamma \equiv \hbar/2\tas$, of the momentum eigenstates
(i.e. the band states defined by a momentum $\hbar \vk$ with $\vk$ as
the conserved wave vector).
The difference between $\tat$ and $\tas$ arises from the subtle effect
of the wave vector dependent impurity potential $u_i(\vk)$ which
distinguishes between momentum scattering in {\it all} directions
contributing to $\tas^{-1}$ and transport relaxation $\tat^{-1}$ which
is unaffected by forward scattering (or more generally, small-angle
scattering) \cite{kn:Dassarma85}.

In this paper we study theoretically
$\tat$ and $\tas$ in 
the 2D graphene layer due to long-range and short-range impurity
scattering, finding interesting behavior in the ratio $\tr$
as a function of impurity location, carrier density, and the system
environment (e.g. the background lattice dielectric constant of the
substrate). We compare graphene $\tr$ with the corresponding situation
in the extensively studied 2D semiconductor systems, finding
significant difference.

In 3D metals or semiconductors, the effective impurity-electron
interaction potential is almost always short-ranged. This is true even
when the bare impurity potential is long-range Coulombic 
%due to random charged centers 
since carrier-induced electronic screening of the
charged impurities is highly effective. This strongly screened
short-range impurity potential in 3D systems (as well as in 2D systems
when electrons and impurities are not spatially separated) has led to
the almost universal theoretical adoption of the uncorrelated
white-noise zero-range impurity potential model for studying the impurity
scattering effect on ``metallic'' transport properties. Since such a
white-noise disorder, by definition, leads to isotropic impurity
scattering dominated entirely by large-angle scattering, $\tat = \tas$
in ordinary metals and semiconductors for such a zero-range disorder
model as forward scattering simply plays no special role
here. Graphene 
is, however, qualitatively different even for this 
short-range white noise disorder model due to its chiral sublattice
symmetry which suppresses backward (i.e. a scattering induced
wave vector change by $2k_F$ from $+k_F$ to $-k_F$) scattering
\cite{kn:dassarma2007a}, thus 
introducing an intrinsic chiral preference for forward scattering over
backward scattering. As we show later in this paper, due to this
suppression of backward scattering, the zero-range white-noise
disorder potential model leads to $\tat = 2\tas$ in graphene in
contrast to the $\tat = \tas$ in ordinary metals and semiconductors.
This is due to the importance of $k_F$ (rather than $2k_F$) scattering
in dominating graphene transport properties whereas the small-angle
scattering always dominates $\tas$.

For long-range impurity scattering, however, it has been shown
\cite{kn:Dassarma85} that $\tat$ can exceed $\tas$ by a large factor,
and in high-mobility 2D modulation-doped semiconductor structures,
where the predominant scattering is the small-angle scattering due to
charged impurities located far from the 2D electron layer, in general
$\tat \gg \tas$. In this paper, we theoretically calculate $\tr$ ( as
well as $\tat$ and $\tas$ individually) for 2D graphene electrons (and
holes) due to both long-range and short-range impurity scattering, taking
into account the dependence on the impurity location. We find that
there are significant differences between 2D graphene (with its linear
chiral dispersion) and 2D semiconductors, showing that $\tr$ in
general could also be very large for graphene although the behavior is
quite different from the corresponding 2D semiconductors.

Our work is restricted entirely to extrinsic or doped (or gated)
graphene where the Fermi level is away from the charge neutral Dirac
point (taken to be the energy zero). There has been extensive recent
theoretical work \cite{Ando2006,Falko,Nomura,Hwang,Adam}
on the charged impurity scattering limited transport properties in
gated extrinsic graphene, but the single-particle quantum relaxation
time $\tas$ (and its relation to the transport scattering time $\tat$)
has not been discussed. We note that while $\tat$ has been
experimentally studied recently in extrinsic graphene \cite{Kim,Fuhrer}
through measurements of density-dependent conductivity, the
corresponding $\tas$ has not yet been directly experimentally
measured. The most direct way of measuring $\tas$ experimentally is by
measuring the quantum level broadening $\Gamma = \hbar/2\tas$ through
the Dingle temperature $T_D = \Gamma/\pi k_B$. The Dingle temperature
can be measured in low-field SdH magnetoresistance oscillations as has
been done extensively in 2D semiconductor structures \cite{SdH}.
Any measurement of the single-particle level broadening,
e.g. magnetization, also gives an estimate for $\tas$. Experimental
comparisons of $\tr$ for various system parameters (e.g. carrier
density, impurity location, materials parameters such as effective
mass and lattice dielectric constant) have been carried out in
semiconductor based 2D systems, but not in graphene since measurements
of $\tas$ are unavailable in graphene. Extensive measurements of
$\tat$ as a function of carrier density, i.e. gate voltage,
however, have been carried out \cite{Kim,Fuhrer} in graphene recently.

In our theoretical calculations (and in the numerical results) we have
used two complementary models of bare disorder: Long-range random
charged impurity centers at or near the graphene-substrate interface
and short-range model white noise disorder in the graphene layer. 
%For charged impurity centers (``Coulomb disorder''), 
We also include screening
(by the graphene carriers themselves) within random phase
approximation (RPA) in the theory. In Sec. II we
describe theoretical details and present our results. In Sec. III we
discuss the relevance and significance of our theory in the context of
transport experiments in gated or doped graphene, and provide a
conclusion.

\section{theory and results}

In the relaxation time approximation 
the graphene transport scattering time $\tau_t$ by
randomly distributed impurity centers is given by
\cite{Hwang,Ando2006} 
\begin{widetext} 
\begin{equation}
\label{eq:scattime}
\frac{1}{\tau_t}  =  \frac{2\pi n_i}{\hbar}\sum_{\lambda'}
\int\frac{d^2k'}{(2\pi)^2} \frac
{<\left |{V_{ei}}(q)\right |^2>}{\varepsilon(q)^2} F_{\lambda
  \lambda'}({\bf k,k}') 
(1-\cos\theta_{kk'})\delta\left ( 
\epsilon_{\lambda{\bf k}} - \epsilon_{\lambda'{\bf k'}} \right ),
\end{equation}
\end{widetext} 
where $n_i$ is the concentration of the 
impurity center, 
$q = |{\bf k} - {\bf k}'|$, $\theta_{{\bf kk}'}$ is the scattering
angle between the scattering in- 
and out- wave vectors ${\bf k}$ and ${\bf k}'$, 
$\lambda,\lambda'=\pm 1$ denote the band indices, 
$\epsilon_{\lambda\bf k} =
\lambda \gamma |{\bf k}|$ is a single particle energy 
($\gamma = \hbar v_F$
being the graphene band velocity), 
and $F_{\lambda \lambda'}({\bf k},{\bf k}')$ is
the overlap of states and given by
$F_{\lambda \lambda'}({\bf k},{\bf k}') = (1 + \lambda \lambda'
\cos\theta_{{\bf k,k}'})/2$.
In Eq. (\ref{eq:scattime}) $V_{ei}(q)$ is the matrix elements
of the scattering potential between an
electron and an impurity. 
For charged impurities, we use the Coulomb interaction
 $V_{ei}(q) = 2 \pi e^2 /(\kappa q)$, where $\kappa$ is the dielectric
 constant of surrounding materials, and for short-range point
defect scatterers,  
$V_{ei}(q) = v_0$, a constant.  
In Eq.~\ref{eq:scattime},
$\varepsilon(q)$ is the static RPA
dielectric (screening) function appropriate for 
graphene~\cite{kn:Hwang2006b}, given by
\begin{equation}
\varepsilon(q) = 1 + v_c(q) \Pi(q),
%\varepsilon(q) = 1 + \frac{q_{TF}}{q} \Pi(q)
\label{die}
\end{equation}
where $v_c(q) = 2 \pi e^2 /\kappa q$ is the electron-electron Coulomb
potential, and $\Pi(q)$
is the polarizability function of graphene which is calculated to be
%\begin{widetext} 
\begin{eqnarray}
\Pi(q) = \left \{ 
 \begin{array}{ll} 1  & \mbox{if $q \leq 2 k_F$} \\
                  1 +\frac{\pi q}{8k_F} - \frac{\sqrt{q^2-
                    4 k_F^2}} {2q}
		  - \frac{q \sin^{-1}(2 k_F/q)}{4k_F}
                  & \mbox{if $q > 2 k_F$} 
\end{array} 
\right .  .
\end{eqnarray}
%\end{widetext}
Since we consider elastic scattering we can 
neglect interband scattering processes ($\lambda' \neq
\lambda$). Then, we have the leading-order (in impurity disorder
$n_i$) transport scattering time
\begin{equation}
\label{eq:scattime_r}
\frac{1}{\tau_t}  =  \frac{n_i}{2\pi\hbar}\frac{E_F}{\gamma^2}
\int_0^{\pi}d\theta \frac
{<\left |{V_{ei}}(q)\right |^2>}{\varepsilon(q)^2} 
(1-\cos^2\theta),
\end{equation}
where $q = 2k_F\sin(\theta/2)$.

From a many-body-theory viewpoint the single particle relaxation time
$\tau_s$ can be calculated from the electron self energy of the  
coupled electron-impurity system \cite{kn:ando1982,Dassarma81}. The
electron self-energy 
due to the impurity scattering is given by
\begin{equation}
\Sigma_{\lambda}(k,\omega)=\sum_{\lambda' {\bf k}'}
\frac{<|V_{ei}(q)^2|>}{\epsilon(q)^2}F_{\lambda \lambda'}({\bf k,k}')
G_{\lambda}({\bf k}',\omega),
\label{sigma}
\end{equation}
where $G_{\lambda}({\bf k},\omega) = 1/(\omega-\epsilon_{\lambda {\bf
    k}} + i \delta)$ is the noninteraction Green's function.
The single particle relaxation time is related to the imaginary part
of the single particle self-energy function by
\begin{equation}
\frac{1}{\tau_s} =\frac{2}{\hbar} {\rm Im} \Sigma(k_F,E_F),
\end{equation} 
with the single particle quantum (impurity induced) level broadening
$\Gamma_s = {\rm Im}\Sigma(k_F,E_F)$, i.e. $\tas = \hbar/2\Gamma_s$.
In the leading order disorder approximation,
we have the single particle relaxation time $\tau_s$
\begin{equation}
\frac{1}{\tau_s}  =  \frac{n_i}{2\pi\hbar} 
\frac{E_F}{\gamma^2}\int_0^{\pi}d\theta \frac
{<\left |{V_{ei}(q)}\right |^2>}{\varepsilon(q)^2} 
(1+\cos\theta).
\label{scat_s}
\end{equation}
Thus, the only difference between the scattering time $\tat$
[Eq. (\ref{eq:scattime_r})] and the
single-particle relaxation time $\tas$ [Eq. (\ref{scat_s})] is the weighting
factor $1-\cos 
\theta$ in the transport scattering time. 
The factor $1-\cos\theta$ weights the amount of backward scattering of the
electron by the impurity. Small angle (or forward) scattering, where
$\cos\theta \approx 1$, is relatively unimportant in contributing to
$\tau_t^{-1}$ and contributes little to the resistivity. In normal 2D
systems the factor $1-\cos\theta$
obviously favors large angle scattering events, which are more
important for the electrical resistivity. However, in graphene the large
angle scattering is also suppressed due to the wave function overlap
factor $1+\cos\theta$.  
The transport scattering time thus gets weighted by an angular
contribution factor of $(1 - \cos \theta)(1+\cos\theta)$, which
suppresses both small-angle 
scattering and large-angle
scattering contributions in the transport scattering rate. However, the
single particle relaxation time is weighted only by $1+\cos\theta$ term.
Therefore, $\tau_t$ is
insensitive to both small and large angle scatterings
while $\tau_s$ is only sensitive to small angle scattering events. 
In fact, the dominant contribution to $\tat$ comes from $\cos^2 \theta
= 0$, i.e. $\theta = \pi/2$ scattering, which is equivalent to $|\vk -
\vk'| \equiv k_F$ ``right-angle''scattering in contrast to the $2k_F$
back scattering in ordinary 2D systems.
This leads to significant difference between graphene and parabolic 2D
semiconductor systems with respect to the behavior of $\tr$.

%We emphasize that both $\tau_t$ and $\tau_s$ can be measured in
%experiments and in general, the two scattering times are quite
%different. This difference arises because the transport scattering
%time is dominated by large-angle scattering contributions, while all
%scattering events contribute with equal weight to the single particle
%relaxation time.  In graphene, the transport scattering time is
%dominated by scattering angle $\theta_{\vk \vk'} = \pi$, which
%corresponds to a momentum transfer of $q = |\vk' - \vk| = k_F$, where
%$k_F$ is the Fermi momentum.

Using RPA screening function [Eq. (\ref{die})] at $T=0$ we calculate 
the scattering times due to charged impurities distributed completely 
randomly at the interface between graphene and the substrate
with density $n_{ic}$
\begin{subequations}
\begin{eqnarray}
\frac{1}{\tau_{tc}} =  \frac{r_s^2}{\tau_{0c}}I_{tc}(2r_s),  \\
\frac{1}{\tau_{sc}} =  \frac{r_s^2}{\tau_{0c}}I_{sc}(2r_s),
\end{eqnarray}
\label{tau_t}
\end{subequations}
where $r_s = e^2/\gamma \kappa$
%, which is corresponding to the
%interaction strength parameter of a normal 2D system (i.e. the ratio of
%potential energy to kinetic energy), 
and
\begin{equation}
\frac{1}{\tau_{0c}} = \frac{2\sqrt{\pi} n_{ic} \gamma}{\hbar
  \sqrt{n}}.
\end{equation}
In Eq. (\ref{tau_t}) $I_{tc}(x)$, $I_{sc}(x)$ are calculated to be
\begin{subequations}
\begin{eqnarray}
I_{tc}(x) & = & \frac{\pi}{2}-2\frac{d}{dx}\left [ x^2g(x) \right ] \\
I_{sc}(x) & = & - \frac{d}{dx} g(x)
\end{eqnarray}
\end{subequations}
where $g(x)$ is given by
\begin{equation}
g(x) = -1 + \frac{\pi}{2}x+(1-x^2)f(x)
\label{gx}
\end{equation}
with
\begin{eqnarray}
f(x) = \left \{ 
 \begin{array}{cl} \frac{1}{\sqrt{1-x^2}} \ln\left[\frac{1+\sqrt{1-x^2}}{x}
   \right ]
                  & \mbox{for  $x < 1$} \\
                   1
                  & \mbox{for $x =1$} \\
                   \frac{1}{\sqrt{x^2-1}}\cos^{-1}\frac{1}{x}
                  & \mbox{for $x > 1$} \end{array} 
\right. .
\label{fx}
\end{eqnarray}
From Eqs. (\ref{gx}) and (\ref{fx}) we have
\begin{equation}
\frac{dg(x)}{dx} =   \frac{\pi}{2} - \frac{1}{x} - 
xf(x).
\end{equation}
%Thus, the functions $I(x)$ can be expressed as a function of  $r_s$  
%\begin{equation}
%I_{tc}(2r_s)  =   \frac{\pi}{2} + 12r_s(1-\pi r_s) + 
%8r_s(6r_s^2-1)f(2r_s) 
%\end{equation}
%\begin{equation}
%I_{sc}(2r_s) = -\frac{\pi}{2} + \frac{1}{2r_s} + 
%2r_sf(2r_s).
%\end{equation}
%where $f(x)$ is given by
The limiting forms of scattering times in the small and large $r_s$ regimes
are given by 
\begin{eqnarray}
\frac{\tau_{0c}}{\tau_{tc}} = \left\{ 
 \begin{array}{ll} r_s^2 \left[{\pi}/2 + 12 r_s+ 8 r_s 
     \ln(r_s) + ... \right ] & \mbox{for  $r_s \ll 1$} \\
{\pi}/{32} - {1}/(15r_s) + O(1/r_s^2)
                  & \mbox{for $r_s \gg 1$} \end{array} 
\right. , 
\end{eqnarray}
and
\begin{eqnarray}
\frac{\tau_{0c}}{\tau_{sc}} = \left\{ 
 \begin{array}{ll}  {r_s}/{2}-{\pi r_s^2}/{2} - 2 r_s^3
     \ln(r_s) + ...  & \mbox{for  $r_s \ll 1$} \\
{\pi}/{16} - {1}/(12r_s) + O(1/r_s^2)
                  & \mbox{for $r_s \gg 1$} \end{array} 
\right. .
\end{eqnarray}

\begin{figure}
\epsfysize=2.4in
\epsffile{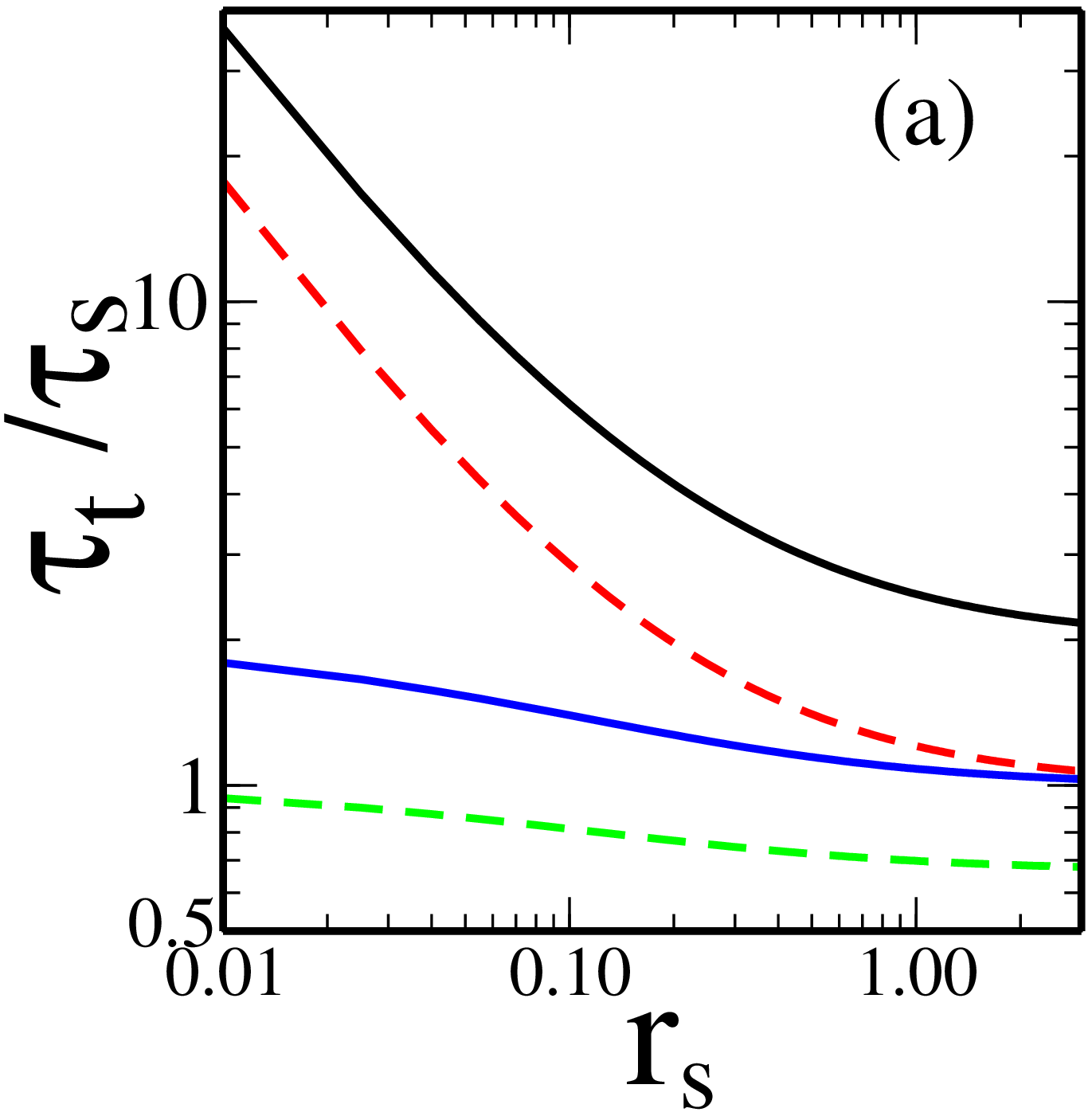}
\epsfysize=2.3in
\epsffile{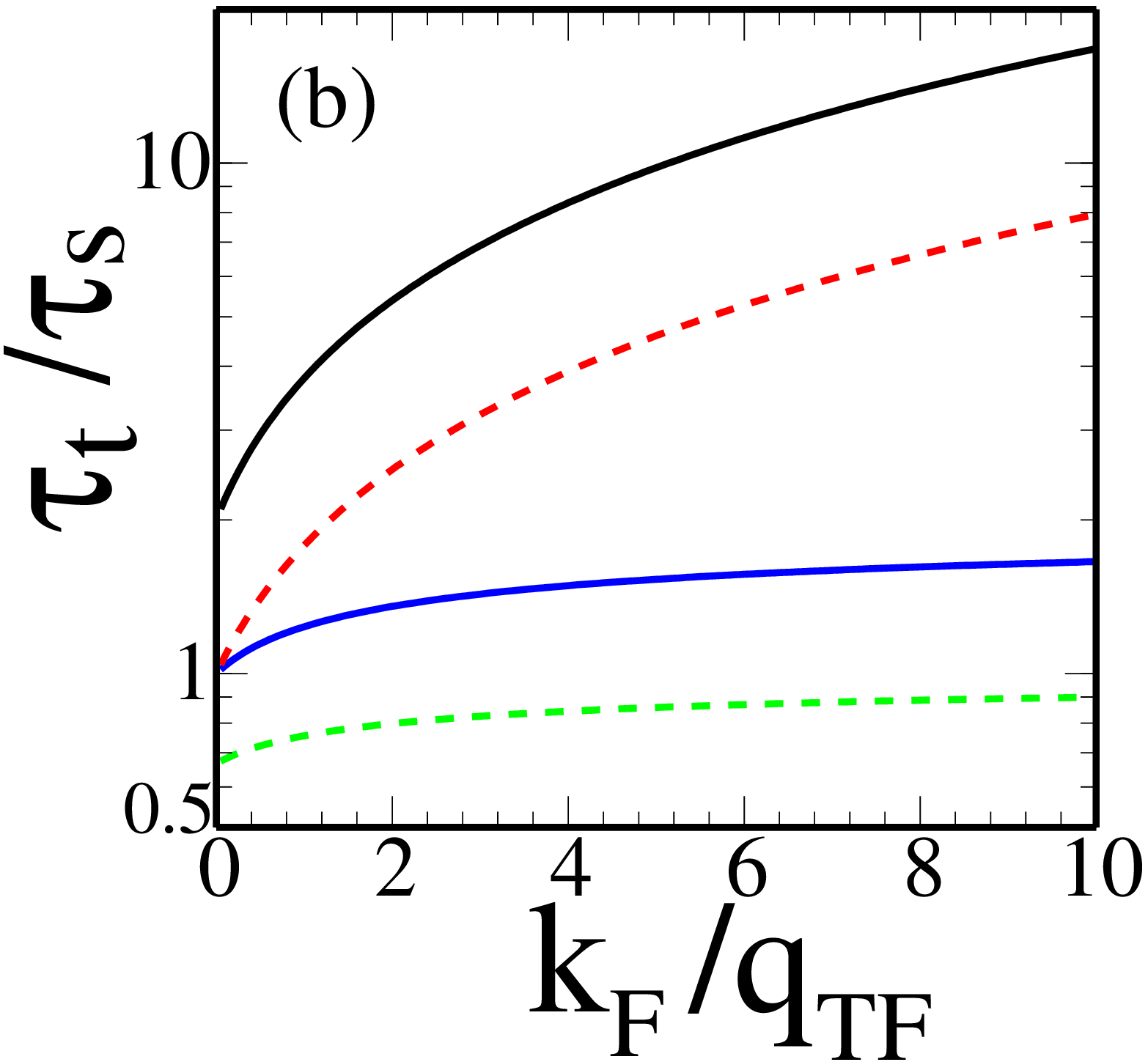}
\caption{\label{Fig1} (Color online)
Calculated the ratio of the transport scattering time to single
particle scattering time $\tau_t/\tau_s$ (a) as a function of the
parameter $r_s=e^2/\gamma \kappa$ and (b) as a function of $k_F/q_{TF}$.
Top (bottom) solid  line
represents the ratio for charged impurity (short-ranged neutral
impurity) scattering.
Top (bottom) dashed line
represents the ratio of normal 2D system with parabolic
dispersion for charged impurity (short-ranged neutral
impurity) scattering. 
We note that for graphene $k_F/q_{TF} \equiv 1/4r_s$.
}
\end{figure}

From Eq. (\ref{tau_t}) we can find easily the ratio of transport
scattering time to the 
single particle relaxation time,
$\tau_{tc}/\tau_{sc}$. Fig. \ref{Fig1} shows the 
ratio as a function of $r_s$. 
Note that the ratio is only a function of $r_s$ and does not
depend on the carrier density.
We have the limiting form of  $\tau_{tc}/\tau_{sc}$, in the small
and large $r_s$ regimes 
\begin{eqnarray}
\frac{\tau_{tc}}{\tau_{sc}} = \left\{ 
 \begin{array}{ll} \frac{1}{\pi r_s} -\frac{16}{\pi^2}\ln(r_s) &
   \mbox{for  $r_s \ll 1$} \\ 
2 + \frac{8}{5\pi}\frac{1}{r_s}
                  & \mbox{for $r_s \gg 1$} \end{array} 
\right. .
\end{eqnarray}
For large $r_s$, the ratio approaches 2, which indicates that
scattering is not isotropic even though the screening is strong.
On the other hand, $\tau_{tc}/\tau_{sc}$ diverges as $1/r_s$ for small $r_s$.
In $r_s \rightarrow 0$ limits, where the screening is very weak, 
the transport scattering rate is much weaker than the single
particle scattering rate, which indicates that the mobility in these
limits is very high.

For short-ranged ($\delta$-function) scatterers with impurity density
$n_{i\delta}$ and potential strength $v_0$ 
we have the scattering times (considering screening effects)
\begin{subequations}
\begin{eqnarray}
\frac{1}{\tau_{t\delta}} = \frac{1}{\tau_{0\delta}} I_{t\delta}(2r_s) \\
\frac{1}{\tau_{s\delta}} = \frac{1}{\tau_{0\delta}} I_{s\delta}(2r_s)
\end{eqnarray}
\end{subequations}
where $1/\tau_{0\delta} = 2n_{i\delta} \sqrt{n}v_0^2/\sqrt{\pi}\hbar \gamma$,
and $I_{t\delta}$, and $I_{s\delta}$ are given by
\begin{subequations}
\begin{eqnarray}
I_{t\delta}(x)& = & \frac{\pi}{8} - \frac{4}{3}x+\frac{3\pi}{2}x^2
 - 2 \frac{d}{dx}\left [ x^4 g(x)\right ] \\ 
I_{s\delta}(x) & = & \frac{\pi}{4}-\frac{d}{dx}\left[ x^2g(x) \right ].
\end{eqnarray}
\end{subequations}
%\begin{eqnarray}
%I_{t\delta}(r_s) & = & \frac{\pi}{16}-\frac{4}{3}r_s + {3\pi}r_s^2
%+ 40r_s^3(1-\pi r_s) \nonumber \\
%               & & -32 r_s^3(1-5r_s^2)f(2r_s)
%\end{eqnarray}
In Fig. \ref{Fig1} we show the ratio of the scattering times for the
short-ranged scatterers, $\tau_{t\delta}/\tau_{s\delta}$, as a
function of $r_s$, and the limiting forms are given by
\begin{eqnarray}
\frac{\tau_{t\delta}}{\tau_{s\delta}} = \left\{ 
 \begin{array}{ll} 2 + {32}r_s\ln(r_s)/\pi &
   \mbox{for  $r_s \ll 1$} \\ 
1 + {128}{r_s}/105\pi
                  & \mbox{for $r_s \gg 1$} \end{array} 
\right. .
\end{eqnarray}
We find that  $1 \le \tau_{t\delta}/\tau_{s\delta} \le 2$ (see
Fig. \ref{Fig1}). The ratio does not exceed 2 for short-ranged
scatterers, but 
for charged impurity scattering the ratio is always greater than
2. Thus the ratio of the scattering times may offer the
possibility of determining the relevant scattering mechanisms in
disordered graphene layers.

Now we compare the calculated scattering time ratio of graphene with
the scattering time ratio of a normal 2D
electron system with parabolic energy dispersion. We 
have the scattering time ratio for charged impurity (see Appendix A) 
in a normal 2D layer as 
\begin{eqnarray}
\frac{\tau_{tc}}{\tau_{sc}} = \left\{ 
 \begin{array}{ll} \frac{1}{\pi r_s} -\frac{4}{\pi^2}\ln(r_s) &
   \mbox{for  $r_s \ll 1$} \\ 
1 + \frac{4}{3\pi}\frac{1}{r_s}
                  & \mbox{for $r_s \gg 1$} \end{array} 
\right. ,
\end{eqnarray}
where $r_s = 1/a_B \sqrt{\pi n}$ ($a_B$ is an
effective Bohr radius of the system and $n$ is a carrier
density). Unlike graphene, in
normal 2D systems $r_s$ depends
on the carrier density. In small $r_s$ (or high
density) limit the ratio shows the same leading order behavior as
graphene (i.e. $\tau_t/\tau_s \propto 1/r_s$), but in 
large $r_s$ (low density) limit  $\tau_t/\tau_s \rightarrow
1$. The difference between normal 2D system and graphene in the strong
screening limit can be traced back 
to the suppression of the backscattering in graphene. 
%The spatial separation between the impurities and the carriers (remote
%doping) strongly enhance the ratio as much as orders of magnitude 
%\cite{kn:Dassarma85}. 
%In high-mobility GaAs based 2D systems the ratio is nearly an order of
%magnitude \cite{SdH}. Thus, the remote charged impurity
%scattering is the main scattering mechanism.
We also find the
difference in the scattering time ratio for short-ranged impurity.
The ratio becomes $2/3 \le \tau_{t\delta}/\tau_{s\delta}
\le 1$. (See Fig. \ref{Fig1} and Appendix A.) Thus, the ratio is
always less than 1. In experiment,
the ratio of the transport time to the single particle time has been
found to be smaller than 1 in Si-MOSFET systems.
Thus, in Si-MOSFET system the short-ranged scattering 
(such as interface roughness scattering)  dominates.

%%%%%%%%%%%%%%%%%%%%%%%%%%%%%%%%%%%%%%
%These theoretical results provide a further method to
%experimentally discriminate between short-range and Coulomb scattering
%as to the relevant scattering mechanism in disordered graphene layers.
%Furthermore, we find that the mobility (which is independent of the
%carrier density if we consider only charged impurities) depends on
%both the charged impurity scattering concentration $n_{ic}$ and the
%substrate dielectric constant $\kappa$ through the interaction
%parameter $r_s = e^2/\gamma \kappa$.  We find that while increasing
%the dielectric constant enhances mobility for Coulomb scatterers, it
%also reduces the relative importance of Coulomb scattering and that
%for $r_s \rightarrow 0$, the mobility reaches a limiting value that
%depends only on neutral impurity scattering. 
%%%%%%%%%%%%%%%%%%%%%%%%%%%%%%%%%%%%%%%%%%

%%%%%%%%%%%%%%%%%%%%%%%%%%%%%%%%%%%%%%%%%%%%%%%%%%%%%%%%
%%%%%%%%       Fig. 1 %%%%%%%%%%%%%%%%%%%
%%%%%%%%%%%%%%%%%%%%%%%%%%%%%%%%%%%%%
%%% Figure template for latex
%%%%%%%%%%%%%%%%%%%%%%%%%%%%%%%%%%%%%%
%%%%%%%%%%%%%%%%%%%%%%%%%%%%%%%%%%%%%%

%%%%%%%%%%%%%%%%%%%%%%%%%%%%%%%%%%%%%
%%% Figure 2
%%%%%%%%%%%%%%%%%%%%%%%%%%%%%%%%
\begin{figure}
\epsfysize=2.4in
\epsffile{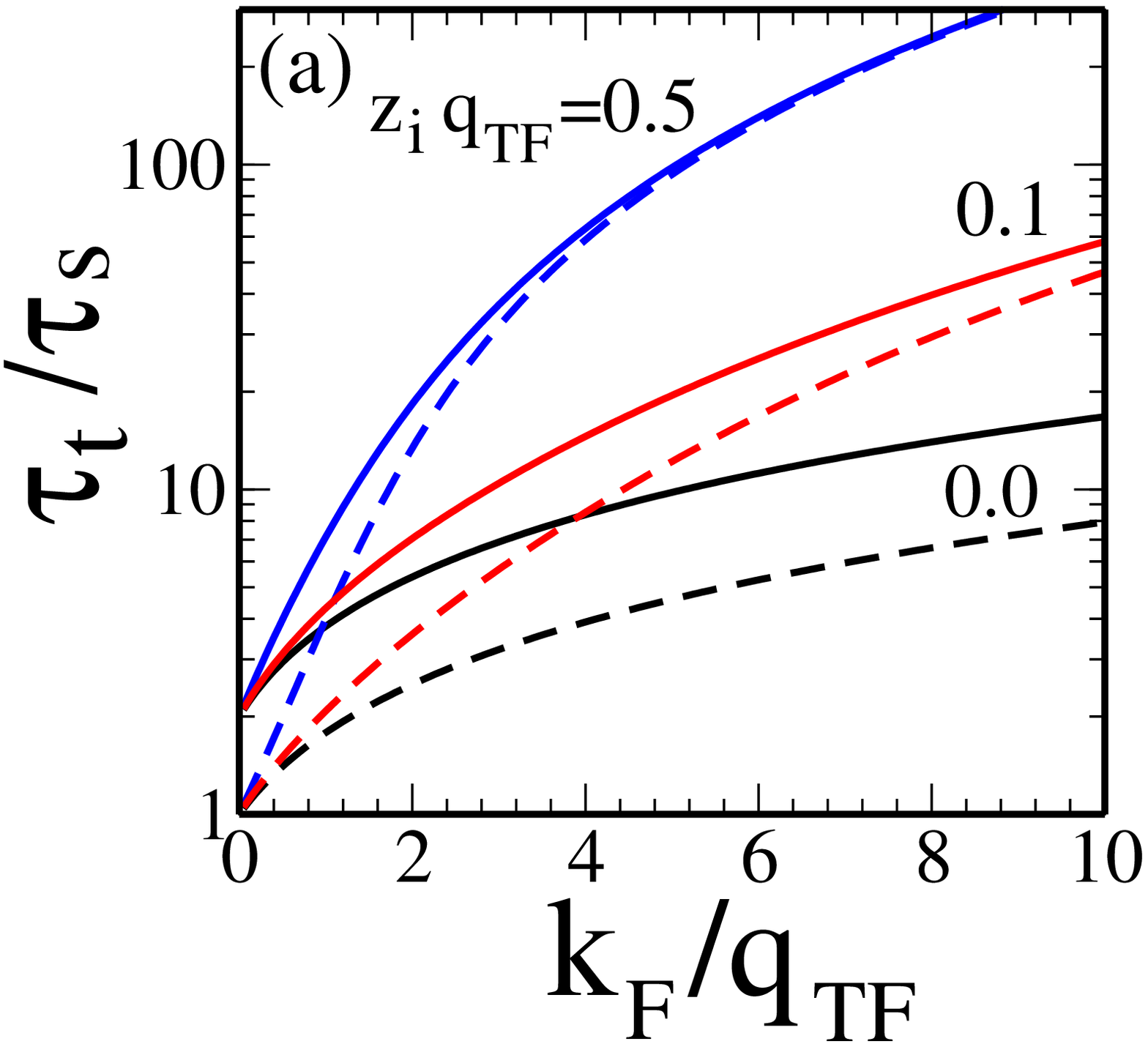}
\epsfysize=2.4in
\epsffile{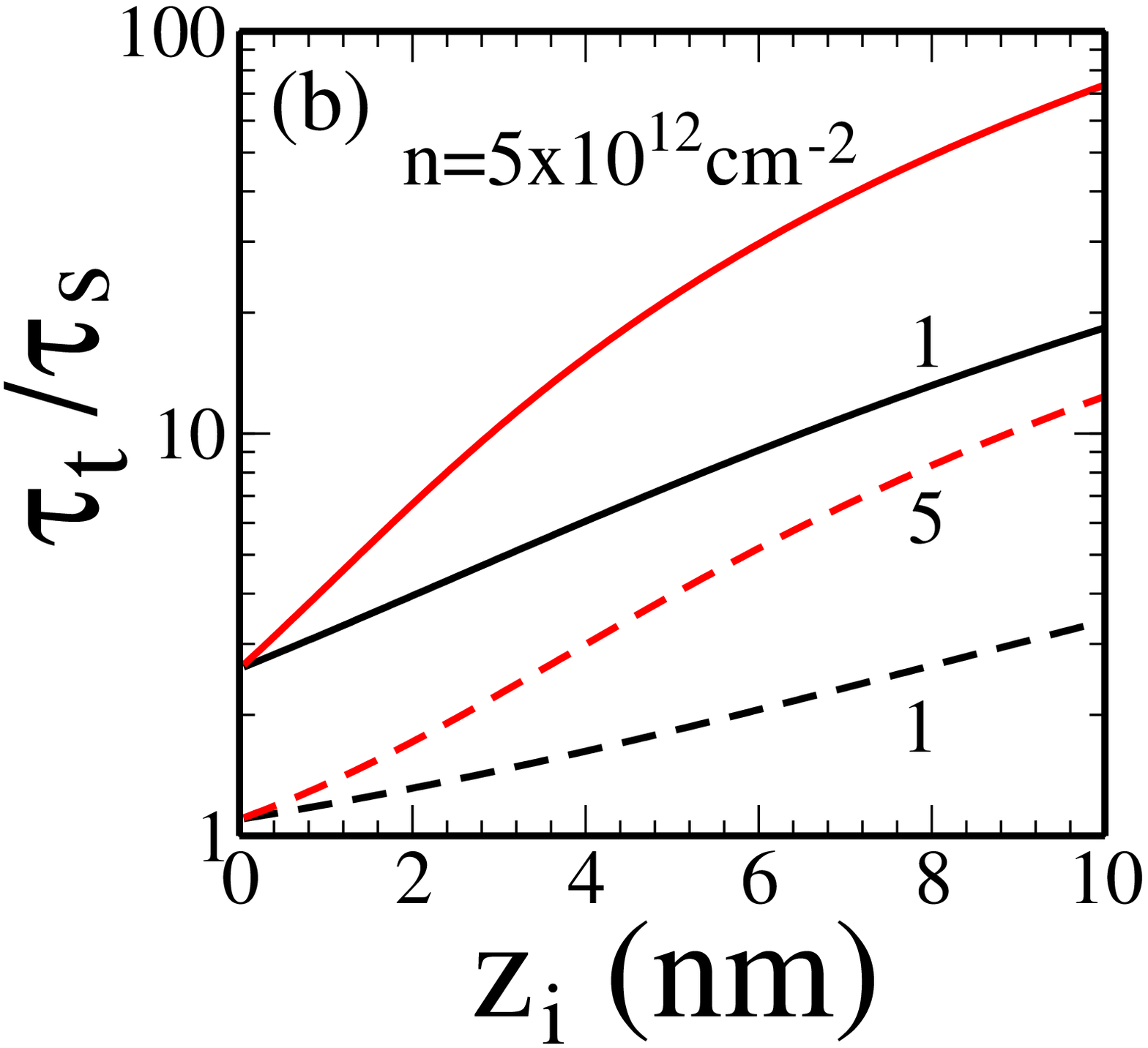}
\caption{ (Color online)
(a) The ration $\tau_t/\tau_s$ of graphene (with substrate
  SiO$_2$, corresponding to $r_s=0.9$) as a function of $k_F/q_{TF}$
for different values of the separation $z_i$ between the
electron layer and the impurity layer. We use the dimensionless
quantity $z_i q_{TF}=0$, 0.1, 0.5. Solid (dashed) lines indicate
results for graphene (normal 2D system). 
(b) The same as Fig. 2(a) as a function of $z_i$ for
  two different electron densities, $n=1,5\times 10^{12}$
  cm$^{-2}$. Solid (dashed) line denote the results 
  for charged Coulomb impurity (short-ranged neutral impurity).
}
\end{figure}

It is, in principle, possible for the charged impurities to be at a
distance ``$z_i$'' away from the 2D graphene layer. In fact, in 2D
modulation doped GaAs-AlGaAs semiconductor heterostructures the
charged dopants are put a distance $d(\equiv z_i)$ away (inside the GaAlAs
insulating barrier region) from the 2D electron gas in order to
minimize the degradation of the electron mobility due to remote dopant
scattering. As discussed in Sec. I of this paper, a large separation
($k_F z_i > 1$) leads to a very strong enhancement of the $\tr$ in the
GaAs-GaAlAs heterostructure since large-angle scattering by the remote
impurities is strongly suppressed by the separation. To see the effect
of separating the impurities from the graphene layer, all we need to
do is to modify the form of the $q$-space Fourier transform of the
electron-impurity Coulomb interaction by the factor $e^{-qz_i}$
arising from the separation between the 2D carriers in the graphene
layer and the charged impurity centers, leading to $V_{ei}(q,z_i) =
V_{ei}(q) e^{-qz_i}$. With this simple modification for $V_{ei}$ we
can calculate the results for $\tat$ and $\tas$ for remote scatterers
using the same formalism as above.

%%%%%%%%%%%%%%%%%%%%%%%%%%%%%%%%%%%%%%%%%%%%%%%
%%%%%%%%%% Fig. 3
%%%%%%%%%%%%%%%%%%%%%%%%%%%%%%%%%%%%
\begin{figure}
\epsfysize=2.4in
\epsffile{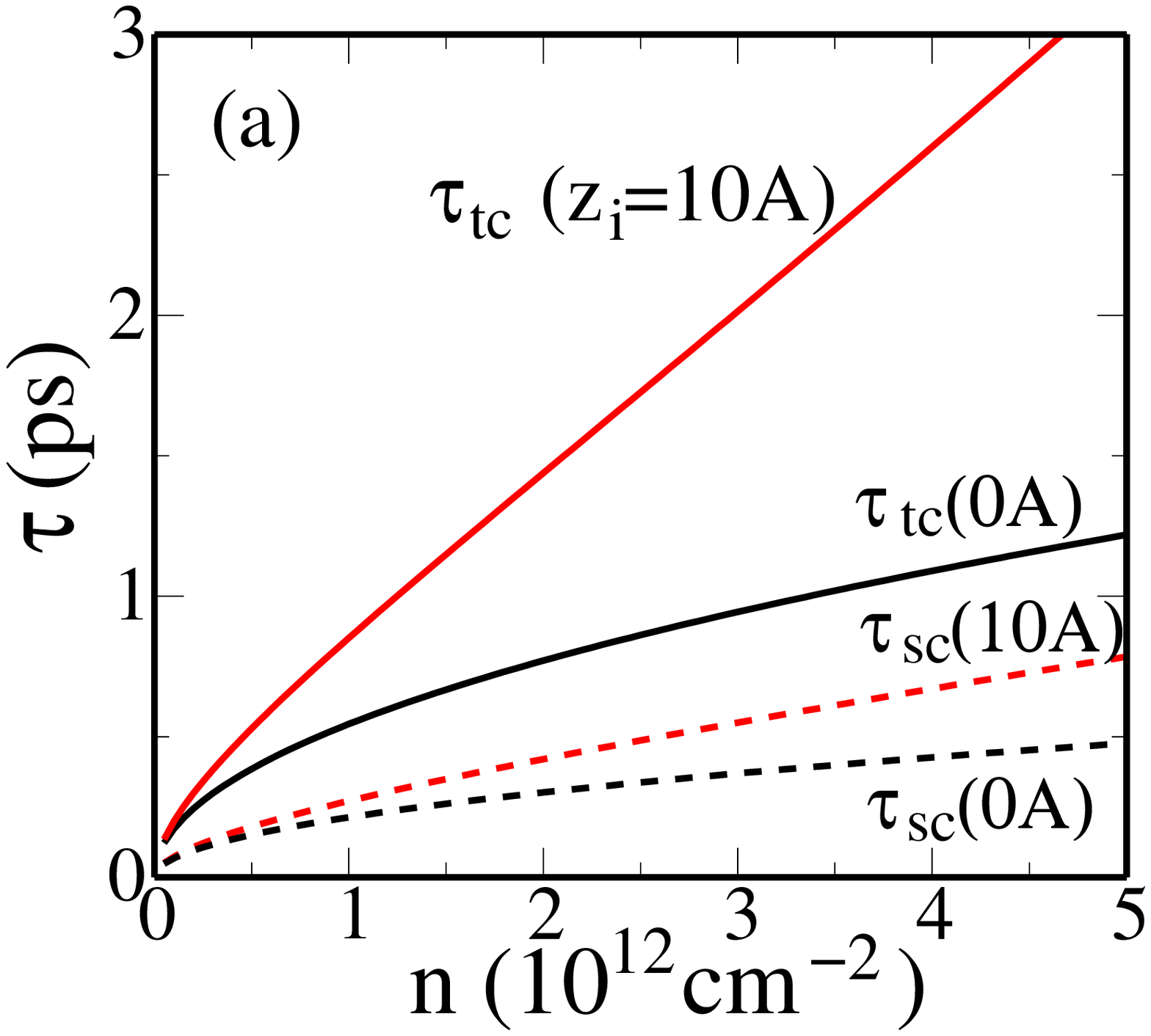}
\epsfysize=2.4in
\epsffile{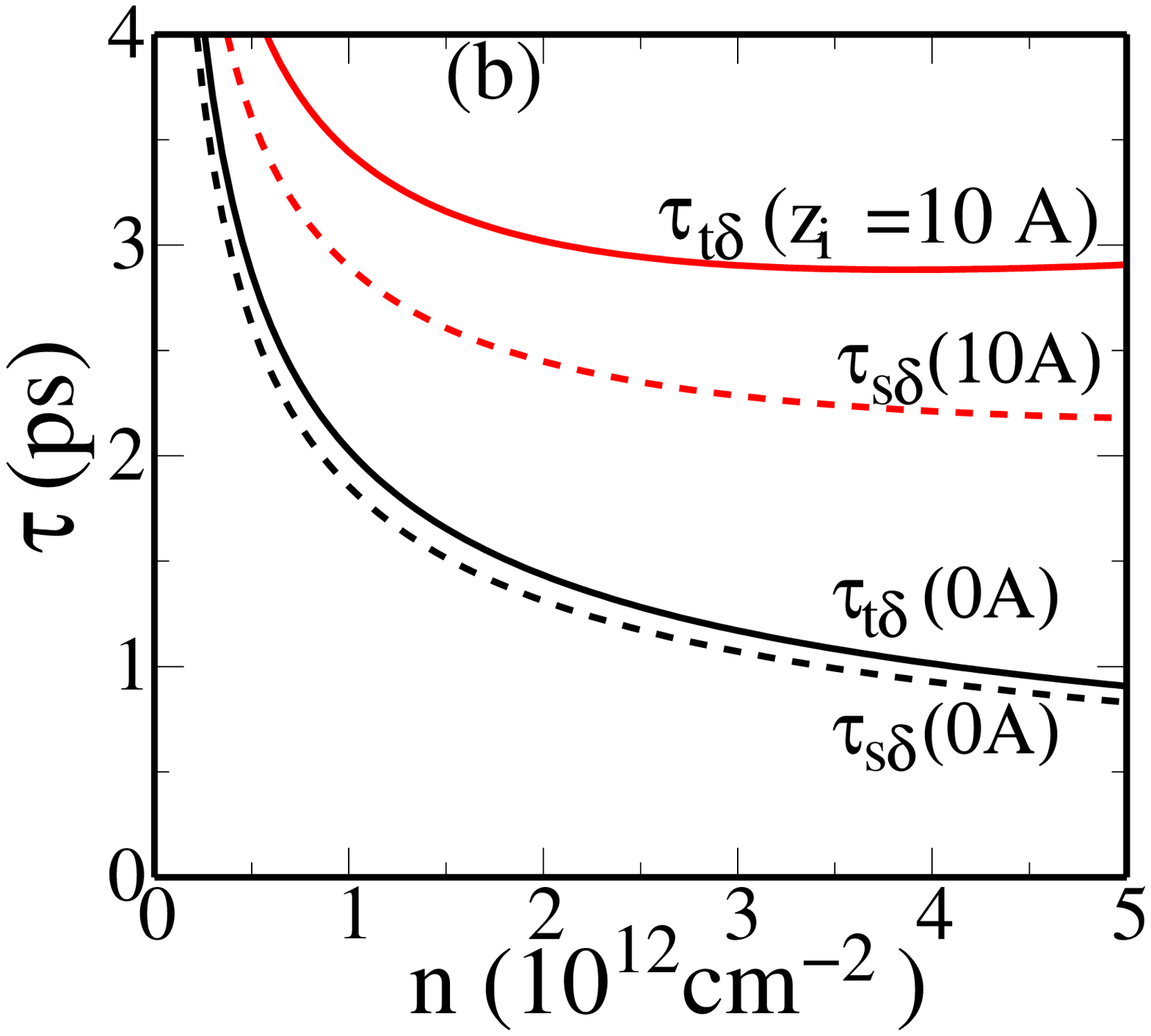}
\caption{ (Color online)
(a) Scattering times (in ps) of graphene (with substrate
  SiO$_2$) as a function of electron density for two different
  impurity locations $z_i=0,10$\AA.
$\tau_{tc}$ ($\tau_{sc}$) indicates the transport scattering time
(single particle relaxation time) for Coulomb impurity scattering. We use an
impurity density $n_{ic}=10^{12}cm^{-2}$. 
The same as Fig. 3(a) for a neutral impurity scattering. We use an impurity
  density $n_{i\delta}=2\times 10^{11}cm^{-2}$ and $v_0=1KeV\AA^2$.
}
\end{figure}

In Figs. 1--4 we present our theoretical results for $\tr$ as well as
$\tat$ and $\tas$ individually in graphene comparing it with the
corresponding regular 2D results (i.e. non-chiral 2D electron system
with parabolic band dispersion). 
We use exactly the same materials parameters for
both graphene and the regular parabolic 2D system. The results are
plotted as a function of the dimensionless parameter $r_s$ (or
$k_F/q_{TF}$), the impurity location from the interface $z_i$, or the
density $n$  where $k_F$ ($q_{TF}$) are 
respectively the Fermi 
wave vector (Thomas-Fermi wave vector) for the system. We give the
expressions for $k_F$ and $q_{TF}$ (with $g$ as the spin and valley
degeneracy factor): $k_F = (4\pi n/g)^{1/2}$ for both graphene and
parabolic 2D systems, $q_{TF} = ge^2k_F/(\kappa v_F)$ for 
graphene and
$q_{TF} = g me^2/(\kappa \hbar^2)$ for 2D parabolic systems.
We note that $q_{TF}/k_F =r_s$ for $g=1$ and $q_{TF}/k_F \propto r_s$
for any value of $g$ for both graphene and parabolic 2D systems.
(In general, $g=4$ for graphene.)

The most important qualitative feature of Figs. 1--2 is that in
general $\tr$ is larger in graphene than in the corresponding 2D
parabolic system for all values of $z_i$ including $z_i=0$ (i.e. when
the impurities are right at the substrate-2D layer interface). For
$z_i \approx 0$ and for $r_s \approx 1$, the situation of current
interest in graphene where the impurities are thought to be within $\sim
1$ nm of the interface (and the substrate is usually SiO$_2$), we
get $\tr \sim 2$ in graphene although it is expected to increase fast
with increasing $z_i$. We therefore suggest a measurement of $\tr$ as
a spectroscopic tool for the determination of the location of the
impurity centers in graphene.
It is interesting to note that in general $(\tr)_{graphene} >
(\tr)_{2D}$ with the difference between them increasing sharply with
increasing $k_F/q_{TF}$ or $z_i$.

In Fig. 3 we show our calculated graphene transport and single
particle times individually as a function of carrier density for
various values of the impurity location $z_i$. In Fig. 3(a) where charged
impurity Coulomb scattering is involved, both $\tat$ and $\tas$
increase with carrier density, but $\tat$ increases much faster,
making $\tr$ increase with increasing density. But for short-range
scattering (Fig. 3(b)), the reverse is true with both $\tat$ and $\tas$
decreasing with increasing density, but $\tas$ decreasing faster,
again leading to an increasing $\tr$ with increasing carrier density.

%%% %%%%%%%%%%%%%%%%%%%%%%%%%%%%%%%%%%%%
%%%%%Figure 4
%%%%%%%%%%%%%%%%%%%%%%%%%%%%%%%%%%%%%
\begin{figure}
\epsfysize=2.4in
\epsffile{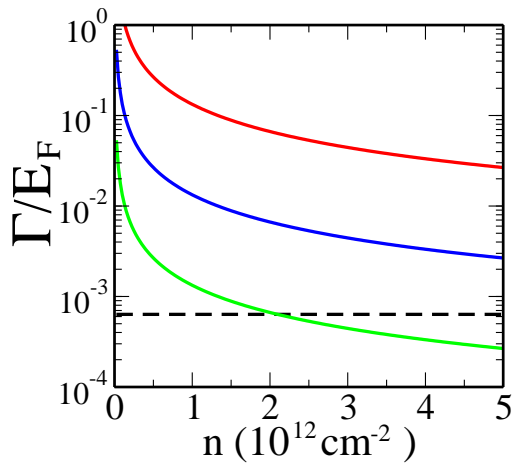}
\caption{ (Color online)
Calculated damping rates scaled by Fermi energy $\Gamma/E_F$
as a function of carrier density for $r_s=0.9$ which corresponds to
graphene sample with SiO$_2$ substrate.  
Solid lines indicate damping rates due to charged impurity scattering with
impurity density $n_{ic} = 100, 10, 1 \times 10^{10}$
  cm$^{-2}$ (from top to bottom), respectively.
Dashed line indicate a damping rate due to short-ranged impurity with
impurity density $n_{i\delta}=10^{10}$ cm$^{-2}$ and potential
strength $v_0=$1 KeV \AA$^2$. We define $\Gamma = \hbar/2\tau_s$.}
\label{damping}
\end{figure}

Finally, in Fig. 4 we show our calculated level broadening $\Gamma =
\hbar/2\tas$ as a function of carrier density in graphene. We note
that for charged impurity scattering $\Gamma/E_F$ decreases
monotonically with increasing carrier density.
However, the scaled damping rate due to
short-ranged impurity scattering is independent of the carrier density.

\section{discussion and conclusion}

We calculate the transport scattering time ($\tau_t$) and the single
particle relaxation time ($\tau_s$) for disordered graphene in the
lowest order of the 
electron-impurity interaction (Born approximation). 
The scattering mechanisms which we consider are screened charged impurity
scattering and 
short-range potential (e.g. caused by lattice defects).  For the 
screening function we use the RPA.
We find that for short ranged scatterers, the ratio of the scattering 
time to the single particle relaxation time is always smaller than
2, but for charged impurity scattering, the ratio is always greater than
2. These theoretical results provide a technique to
experimentally discriminate between short-range and Coulomb scattering
as to the relevant scattering mechanism in disordered graphene layers.
We also find a strong dependence of the ratio $\tr$ in graphene on
$z_i$ the separation of the impurities from the 2D graphene
layer. Somewhat surprisingly the dependence of $\tr$ on $z_i$ is
stranger in graphene than in the corresponding parabolic 2D system,
leading to the possibility that an accurate measurement of $\tr$ in
graphene could lead to better understanding of the impurity location
underlying graphene disorder. In particular, $\tr$ increases rapidly
with increasing $z_i$ in graphene (similar to non-chiral 2D GaAs-GaAlAs
modulation doped heterostructure) and this could be directly
experimentally tested.

We conclude  by emphasizing that independent measurements of $\tat$
and $\tas$ in graphene samples could lead to detailed useful insight
into the nature of disorder scattering of graphene carriers.

\begin{acknowledgements}

This work is supported by U.S. ONR, NSF-NRI, and SWAN.

\end{acknowledgements}

\appendix
\section{}
Here we provide the scattering times of a normal 2D system with
parabolic band. 
Using 2D RPA screening function \cite{kn:ando1982} at $T=0$ we have 
the scattering times for charged impurity centers
\begin{subequations}
\begin{eqnarray}
\frac{1}{\tau_{tc}} =  \frac{1}{\tau_{0c}}I_{tc}(q_0), \\
\frac{1}{\tau_{sc}} =  \frac{1}{\tau_{0c}}I_{sc}(q_0),
\end{eqnarray}
\end{subequations}
where 
\begin{equation}
\frac{1}{\tau_{0c}} = {2\pi\hbar} \frac{n_{ic}}{m} (\frac{2}{g})^2q_0^2,
\end{equation}
$n_{ic}$ is the density of charged impurity, $q_0 = q_{TF}/2k_F$ ($q_{TF} =
g/a_B$ is a 2D Thomas-Fermi wave 
vector with effective Bohr radius $a_B = \hbar^2/me^2$), and
$I_{tc}(q_0)$, $I_{sc}(q_0)$ are given by 
\begin{equation}
I_{tc}(q_0) = \pi-2\frac{d}{dq_0}\left[q_0^2f(q_0)\right ]
\end{equation}
\begin{equation}
I_{sc}(q_0) =  -\frac{d}{dq_0}f(q_0)
\end{equation}
where $f(x)$ is given in Eq. (\ref{fx})
and its derivative is given by
\begin{equation}
\frac{d f}{dx} = \frac{1}{x}\frac{1}{x^2-1}\left[1-x^2f(x) \right]
\end{equation}

For short-ranged impurity centers we have
\begin{subequations}
\begin{eqnarray}
\frac{1}{\tau_{t\delta}} =  \frac{1}{\tau_{0\delta}}I_{t\delta}(q_0), \\
\frac{1}{\tau_{s\delta}} =  \frac{1}{\tau_{0\delta}}I_{s\delta}(q_0),
\end{eqnarray}
\end{subequations}
where 
\begin{equation}
\frac{1}{\tau_{0\delta}} = \frac{2n_{i\delta}}{\pi\hbar^3} mv_0^2,
\end{equation}
$n_{i\delta}$ is the density of short-ranged impurity,
$n_{i\delta}$, $v_0$ is the potential strength and
$I_{t\delta}(q_0)$, $I_{s\delta}(q_0)$ are given by 
\begin{equation}
I_{t\delta}(q_0) = \frac{\pi}{2}-4q_0+3\pi
q_0^2-2\frac{d}{dq_0}\left[q_0^4f(q_0)\right ] 
\end{equation}
\begin{equation}
I_{sc}(q_0) =  \frac{\pi}{2}-\frac{d}{dq_0}\left [ q_0^2f(q_0) \right ].
\end{equation}

\end{document}